\documentstyle[12pt]{article}
\newcommand{\pabar}{\not{\!\partial}}
\newcommand{\Od}{{\cal O}}

\newcommand{\tr}{\mbox{tr}}

\newcommand{\Dbar}{\not{\!{\!D}}}

\newcommand{\mapright}[1]{\smash{\mathop{\hbox to 1cm{\rightarrowfill}}\limits_{#1}}}

\def\gappeq{\mathrel{\rlap {\raise.5ex\hbox{$>$}}
{\lower.5ex\hbox{$\sim$}}}}

\def\lappeq{\mathrel{\rlap{\raise.5ex\hbox{$<$}}
{\lower.5ex\hbox{$\sim$}}}}

\begin{document}
\input epsf \renewcommand{\topfraction}{0.8}
\pagestyle{empty}
\begin{flushright}
\end{flushright}
\vspace*{5mm}
\begin{center}
\Large{\bf Casimir effect between moving branes} \\
\vspace*{1cm} 
\large{\bf Antonio L. Maroto}$^{*}$  \\
\vspace{0.3cm}
\normalsize 
Physics Department, Stanford University, \\
Stanford CA 94305-4060, USA \\
and\\
Departamento de F\'{\i}sica Te\'orica\\
Universidad Complutense de Madrid\\
28040 Madrid, Spain

\vspace*{1cm}  
{\bf ABSTRACT} \\ \end{center}
We consider a  supersymmetric model  with a
single matter supermultiplet in a five-dimensional
space-time with orbifold compactification
along the fifth dimension. The boundary conditions
on the two orbifold planes are chosen in such a way 
that supersymmetry remains unbroken on the boundaries. 
We calculate the vacuum energy-momentum
tensor in a configuration in which the boundary branes are moving with
constant velocity. The results show that  the contribution
from fermions cancels that of bosons only in the static limit, 
but in general a velocity-dependent Casimir energy arises
between the branes. We relate this effect to the particle
production due to the branes motion and finally we discuss some 
cosmological consequences.

\vspace*{5mm}
\noindent

\vspace*{0.5cm}

\noindent PACS numbers: 11.10.Kk, 98.80.Cq 

\vspace{4cm}

\noindent 
\rule[.1in]{8cm}{.002in}

\noindent $^*$E-mail: maroto@fis.ucm.es
\vfill\eject

\setcounter{page}{1}
\pagestyle{plain}

\newpage

\section{Introduction}
The presence of non-trivial boundary conditions in quantum field theory 
leads to interesting phenomena such as the shift in the
zero-point energy with respect to the
unbounded space configuration \cite{Casimir}. In the case of electromagnetic
fields between perfectly conducting parallel plates, such  shift  
has been observed as an atractive force acting on the plates (see 
\cite{onofrio} for a recent measurement). 
Since the
original work of Casimir, this effect has been studied in different contexts,
thus for instance, it plays a fundamental role in hadron physics, 
in the so-called bag model
in which quarks and gluons are confined inside a spherical shell and the
corresponding Casimir energy contributes to the total hadron mass. Also
in higher-dimensional theories, the stability of the compactified extra
dimensions depends crucially on this effect. Other applications can be found
in condensed matter physics and atomic or molecular physics (for
a review see \cite{Bordag}). 
  
More recently, the increasing interest in the construction of cosmological 
models
based on string or M theory has focused on the so-called
brane-world scenario, in which our universe is understood as a 3-brane living
in a higher dimensional space-time. A particularly interesting model appears in 
the low-energy regime of M-theory described by eleven dimensional supergravity
\cite{HW}.
After compactification of six dimensions  in a Calabi-Yau manifold,
we are left with a five-dimensional model in which
matter fields live on the two boundary branes arising after 
$S^1/Z_2$-orbifolding
the fifth dimension. In this context, the Casimir energy arising
between the two static boundaries has been computed in \cite{Horawa,Peskin},  
in the first of these two works,  
the backreaction on the geometry was taken into account. The same
problem has been considered in five-dimensional anti-deSitter space in  
\cite{Odintsov}. 
The recently proposed ekpyrotic (cyclic) model of the universe \cite{TS}
is also based on this framework in which the motion and collision of two 
such branes is responsible for the 
Big-Bang of the standard cosmology. 

In this paper we are interested in studying the possible effects of the
Casimir energy in an scenario like the one mentioned before in which two
branes are moving towards each other. The complete analysis of the problem
is in general too involved to obtain explicit analytic results and, for
that reason, we will consider a simplified model in which the two branes
are moving with constant relative velocity and they are perfectly flat,
ignoring possible gravitational effects.
In any realistic model of a brane collision process it will be necessary
to consider the  acceleration and the brane curvature \cite{Rasanen}, 
but the present analysis would be the first (velocity-dependent) 
correction to the flat static case. We will also consider a simple global 
supersymmetric model
in five dimensions with a single matter hypermultiplet instead of working
with the full supergravity lagrangian in eleven dimensions. Despite these
approximations the results will shed  light on the velocity dependence
of the Casimir energy and therefore on the
stability of the branes system. We will be interested also in the role
of supersymmetry in the cancellation of the Casimir energy in the moving
case. For that purpose we will need to extend previous studies 
of the dynamical Casimir
effect \cite{Bordag1} to fermionic fields. 
In fact we will show that only in the static limit
the contribution from fermions cancels that of bosons (as expected from
supersymmetry), but in general
a velocity dependent Casimir energy arises between the branes.

The paper
is organized as follows, in Section 2, we introduce the five-dimensional
supersymmetric model and set the boundary conditions for the different
fields. In Section 3 we compute the scalar contribution to the Casimir
stress tensor using the image method for Green functions for arbitrary
value of the brane velocity. In Section 4, we perform a similar calculation
but with the
Bogolyubov transformations technique in the
low-velocity regime. Section 5 is devoted to the 
fermionic contribution, and finally Section 6 contains a discussion on the
possible cosmological consequences, paying special attention to the
ekpyrotic (cyclic) model of the universe. 

\section{Supersymmetric model Lagrangian in 5-D}

Let us consider the simplest globally supersymmetric model in 
5-D. A five-dimensional off-shell
hypermultiplet consists of  two complex scalar fields, which we denote 
by $\phi_1$ and $\phi_2$, one four-component Dirac spinor $\psi$ and two 
auxiliary complex
scalars fields $F$ and $G$ \cite{West,Peskin,Pomarol}. The smallest 
representation of the
Clifford algebra in five dimensions is four dimensional and is given by:
$\gamma^A=\{\gamma^0,\gamma^1,\gamma^2,\gamma^3,\gamma^4\}$ where 
$\gamma^4=i\gamma^5$ with the usual definition 
$\gamma^5=i\gamma^0\gamma^1\gamma^2\gamma^3$. With this definition we have
$\{\gamma^A,\gamma^B\}=2\eta^{AB}$ where $\eta^{AB}=\mbox{diag}(+,-,-,-,-)$.
As a consequence the
chirallity operator $\gamma^0\gamma^1\gamma^2\gamma^3\gamma^4= {\bf 1}$
is trivial and it is not possible to define chiral fermions in five dimensions.

The simplest supersymmetric action corresponds to massless 
non-interacting fields and is given by:
\begin{eqnarray}
S=\int d^5x \left(\eta^{AB}\partial_A \phi_1^*\partial_B \phi_1 + 
\eta^{AB}\partial_A \phi_2^*\partial_B \phi_2
+i\bar \psi \gamma^A\partial_A \psi+
\vert F\vert^2+\vert G\vert^2\right)
\label{lag}
\end{eqnarray}

As commented in the introduction, we will assume that the fifth dimension
is compactified on the orbifold $S^1/Z_2$, that is, 
the fifth dimension is just an interval $[0,L]$, with two 
parallel boundary planes at $x^4=0,L$. In fact,  in order to perform the
orbifold projection, we need to assign $Z_2$ parities $\eta=\pm 1$ to the
different fields in such a way that $\Phi(x^\mu,x^4)=\eta \Phi(x^\mu,-x^4)$
where $\Phi$ is any of the hypermultiplet components. A consistent assignment
which respects $N=1$ supersymmetry on the boundaries is given in
Table 1 (see \cite{Igarashi,Peskin}), 
where we have used the definitions $\psi_L=P_L\psi=\frac{1-\gamma^5}{2}\psi$ 
and $\psi_R=P_R\psi=\frac{1+\gamma^5}{2}\psi$.

For an $S^1$ compactification with radius $R$, 
the Fourier expansions for the fields with different $Z_2$ parities 
become:
\begin{eqnarray}
\Phi_+(x^\mu,x^4)=\sum_{n=0}^{\infty} \cos(nx^4/R)\Phi_n(x^\mu)\nonumber\\
\Phi_-(x^\mu,x^4)=\sum_{n=1}^{\infty} \sin(nx^4/R)\Phi_n(x^\mu)
\end{eqnarray}
We see that  even-parity fields satisfy Neumann conditions on the
boundary planes, whereas odd-parity
fields satisfy Dirichlet  conditions  i.e. for $L=\pi R$ we have:
\begin{eqnarray}
\partial_4 \Phi_+(x^\mu,x^4)\vert_{x^4=0,L}&=&0, \nonumber \\
\Phi_-(x^\mu,x^4)\vert_{x^4=0,L}&=&0.
\label{boundary}
\end{eqnarray}

\begin{table}[h]
\begin{center}
\begin{tabular}{|c|c|}
\hline
  & \\ 
 $\eta=1$  & $\eta=-1$  \\
(Neumann)  & (Dirichlet)\\
& \\
\hline  $\phi_1$ & $\phi_2$\\ 
\hline $\psi_R$ & $\psi_L$ \\
\hline  $F$ & $G$\\ \hline 
\end{tabular}
\caption{\footnotesize{Parity assignments for the model in (\ref{lag})}}
\end{center}
\end{table} 

If supersymmetry remains unbroken we expect that the contribution
from fermions cancels that of bosons in the vacuum energy-momentum tensor.
As we will see this is the case when the boundaries are static, however
here we will consider a more general situation in which the boundaries are
moving.    
\section{Scalar contribution}
Let us write the classical energy-momentum tensor for massless real 
scalars  in a five-dimensional Minkowski background:
\begin{eqnarray}
T_{AB}^S&=&(1-2\xi)\phi_{,A}\phi_{,B}
+\left(2\xi-\frac{1}{2}\right)g_{AB}
g^{CD}\phi_{,C}\phi_{,D}\nonumber \\
&-&2\xi\phi_{,AB}\phi+\frac{2}{5}\eta_{AB}\phi\Box\phi
\end{eqnarray}
where for generality we have considered arbitrary
 non-minimal couplings
in the scalar lagrangian $\xi R \phi^2$, which vanish in flat space but
however have a non-vanishing contribution to the energy-momentum tensor.
The particular case $\xi=3/16$ corresponds to the conformal coupling
in five dimensions

In order to calculate the vacuum expectation value of the energy-momentum tensor
we will use the Green's function method. With that purpose we introduce the
so called Hadamard function:
\begin{eqnarray}
D^{(1)}(x,y)&=&\langle 0\vert \phi(x)\phi(y)+\phi(y)\phi(x)\vert 0\rangle 
\end{eqnarray}
where $\vert 0\rangle$ denotes the usual Minkowski vacuum
state built out of  the plane-wave solutions of the free equations 
of motion written in cartesian coordinates. It is possible to write the 
vacuum expectation value
of $T_{AB}$ in cartesian coordinates as:
\begin{eqnarray}
\langle 0\vert T_{AB}^S\vert 0\rangle 
= \lim_{x'\rightarrow x,\; y'\rightarrow x}\frac{1}{2}
\left((1-2\xi)\partial_A^{x'}\partial_B^{y'}
+\left(2\xi-\frac{1}{2}\right)
\eta_{AB}
\eta^{CD}\partial_C^{x'}\partial_D^{y'}
-2\xi\partial_A^{x'}\partial_B^{x'}
\right)D^{(1)}(x',y')
\label{uno}
\end{eqnarray}
where we have used $\Box^{x'}D^{(1)}(x',y')=0$. 
In the simplest case, i.e., unbounded Minkowski space,  
the Hadamard function can be obtained from the mode expansion of the quantum
fields and it is given by:
\begin{eqnarray}
D^{(1)}(x,y)&=&\int \frac{d^4p}{2p_0}\left(e^{ip(x-y)}+e^{ip(y-x)}\right)
\nonumber\\ 
&=&
\frac{1}{4\pi^2}\frac{\theta(\sigma(x,y))}{\sigma(x,y)^{3/2}} 
\label{sfree}
\end{eqnarray}
where $\sigma(x,y)=-\eta_{AB} (x^A -y^A)(x^B-y^B)$, $\theta(\sigma(x,y))$
is the step function 
and $d^4p$  is the integration measure for the spatial components of the
five-momentum $p_A$. 

We will first consider the case in which the two branes at $x^4=0,L$
are fixed. The Hadamard functions satisfying Dirchlet and Neumann 
conditions can be calculated by the image method
(see for example \cite{Igarashi}).
According to the principle of mirror reflection
such functions can be obtained  as
a sum over an infinite number of the free functions in (\ref{sfree}) 
evaluated at the
image points $x'_n$.
They can be written  as:
\begin{eqnarray}
D^{(1)}_{D/N}(x,x')=\sum_{n=-\infty}^{\infty}
(D^{(1)}(x,x'_{n+})\mp D^{(1)}(x,x'_{n-}))
\label{Di}
\end{eqnarray}
where the minus sign corresponds to the Dirichlet function and
the plus sign to  the Neumann one. The image points  are given by:
\begin{eqnarray}
x'_{n\pm}&=&(t',x'^1,x'^2,x'^3,\pm x'^4-2nL)
\label{image}
\end{eqnarray}
These functions satisfy the corresponding boundary conditions given
in (\ref{boundary}).
Since in our model, one complex scalar field 
satisfies  Dirichlet conditions and the other one  Neumann conditions,  
the total scalar energy-momentum tensor  
will be given by the sum of the two kinds of contributions, i.e.:
\begin{eqnarray}
\langle 0\vert T_{AB}^S\vert 0\rangle 
= \lim_{x',y'\rightarrow x}2
\left((1-2\xi)\partial_A^{x'}\partial_B^{y'}
-2\xi\partial_A^{x'}\partial_B^{x'}\right.\nonumber \\
+\left.\left(2\xi-\frac{1}{2}\right)\eta_{AB}
\eta^{CD}\partial_C^{x'}\partial_D^{y'}
\right)\sum_{n=-\infty}^{\infty}
D^{(1)}(x',y'_{n+})
\label{scalar}
\end{eqnarray}
The expression in (\ref{scalar}) 
is divergent in the coincidence limit,  the
divergent contribution coming from the $n=0$ term which is nothing but
the unbounded space vacuum energy-momentum tensor. Thus we can
define the renormalized vacuum expectation value substracting this
free space term:
\begin{eqnarray}
\langle 0\vert T_{AB}\vert 0\rangle_{ren}=
\langle 0\vert T_{AB}\vert 0\rangle-
\langle 0\vert T_{AB}\vert 0\rangle_{n=0}   
\label{ren}
\end{eqnarray}
The final result  does not depend on the non-minimal parameter $\xi$
and  is given by the well-known expression:
\begin{eqnarray}
\langle 0\vert  T_{AB}^S\vert 0\rangle_{ren}^{static}=
-\frac{3\;\zeta(5)}{32\pi^2  L^5}
\left(\begin{array}{ccc}
1&0&0\\
0& -{\bf 1} &0 \\
0&0& 4\end{array}
\right)
\label{scalarest}
\end{eqnarray}

Let us consider next the case in which the 3-brane at $x^4=0$
is fixed whereas the other one is moving with constant velocity $v$
in the fifth dimension. In order to obtain the image points it is easier to 
work in a coordinate system in which the branes are fixed, but the
geometry is contracting (expanding) along the fifth dimension \cite{jauregui}. 
Thus, let us
consider the new coordinates $(\tau,x^1,x^2,x^3,\chi)$ defined as:
\begin{eqnarray}
x^0=\tau \cosh(\chi),\,\,\,\,
x^4=\tau \sinh (\chi)
\end{eqnarray}
In these ccordinates the branes positions
are given simply by $\chi=0$ and $\chi=\chi_0$. Notice that curves 
with constant
fifth coordinate $\chi=\chi_0$ describe brane motions with a constant velocity 
$v=\tanh(\chi_0)$. The corresponding metric tensor reads:
\begin{eqnarray}
ds^2=d\tau^2-(dx^1)^2-(dx^2)^2-(dx^3)^2-\tau^2 d\chi^2
\label{Milne}
\end{eqnarray}
which is nothing but the Milne metric in five dimensions. This metric  
can describe the portions of Minkowski space-time with either  $x^0>0$ or
$x^0<0$. We will use indices $A,B,C,...=0,1,2,3,4$ for Minkowski coordinates 
and $M,N,P,...=0,1,2,3,4$ for the Milne coordinates. 
For this metric the only non-vanishing Christoffel
symbols are 
$\hat\Gamma^\tau_{\chi\chi}=a^2\tau$, $\hat\Gamma^\chi_{\tau\chi}=
\hat\Gamma^\chi_{\chi\tau}=\frac{1}{\tau}$. In the following, 
we will use the hat to denote
objects written in Milne coordinates. 
The invariant interval $\sigma(x,x')$ defined before can be 
written in the new coordinates as:
\begin{eqnarray}
\sigma(x,x')&=&-\tau^2-\tau'^2+2\tau\tau' \cosh(\chi-\chi')+(x^1-x'^1)^2
\nonumber \\
&+&(x^2-x'^2)^2+(x^3-x'^3)^2
\end{eqnarray}
The Dirichlet and Neumann functions are still given by (\ref{Di}), 
but with the image points written in the new coordinates, i.e.:
\begin{eqnarray}
x'_{n\pm}&=&(\tau',x'^1,x'^2,x'^3,\pm \chi'-2n\chi_0)
\end{eqnarray}
The boundary conditions satisfied by the Green functions
now read:
\begin{eqnarray}
D^{(1)}_D(x,x')\vert_{\chi,\chi'=0,\chi_0}&=&0,\nonumber \\
\partial_{\chi} D^{(1)}_N(x,x')\vert_{\chi,\chi'=0,\chi_0}&=&0
\end{eqnarray}

Using Eqs. (\ref{uno}), (\ref{ren}) and changing to Milne
coordinates, we obtain the renormalized expressions for the
energy-momentum tensor. In these coordinates 
the tensors are diagonal and do not depend on the spatial coordinates,
however unlike the static case, they do depend on the $\xi$ parameter. 
Thus for minimal coupling $\xi=0$ we get:
\begin{eqnarray}
\langle 0\vert \hat T_{MN}^S\vert 0\rangle_{ren}^{mov}=
-\frac{3}{32\pi^2\vert\tau\vert^5}\sum_{n=1}^{\infty}
\frac{1}{\vert\sinh^5(n\chi_0)\vert}
\left(\begin{array}{ccc}
1&0&0\\
0& -\frac{3\cosh(2n\chi_0)-1}{2}{\bf 1} &0 \\
0&0& 4\tau^2\end{array}
\right)
\end{eqnarray}
whose trace is non-vanishing. Notice that when the branes
are fixed the result (\ref{scalarest}) is traceless regardless of the
value of the $\xi$ parameter. For the conformal 
coupling $\xi=3/16$, we get:
\begin{eqnarray}
\langle 0\vert \hat T_{MN}^S\vert 0\rangle_{ren}^{mov}=
-\frac{3}{32\pi^2\vert\tau\vert^5}\sum_{n=1}^{\infty}
\frac{5+3\cosh(2n\chi_0)}{8\vert\sinh^5(n\chi_0)\vert}
\left(\begin{array}{ccc}
1&0&0\\
0& -{\bf 1} &0 \\
0&0& 4\tau^2\end{array}
\right)
\end{eqnarray}
which is traceless as expected. These results are valid
for arbitrary values of the brane velocity. However, 
in order to extract an explicit velocity 
contribution to the Casimir energy-momentum tensor, we will 
compare them in 
the non-relativisitic limit $v\ll 1$ with the static result in 
(\ref{scalarest}). In Milne
coordinates, the proper distance between the two branes is given by
$d=\tau\chi_0$, thus, in such limit, we get for minimal coupling:
\begin{eqnarray}
\langle 0\vert \hat T_{MN}^S\vert 0\rangle_{ren}^{mov}\simeq
-\frac{3\zeta(5)}{32\pi^2 d^5}
\left(\begin{array}{ccc}
1&0&0\\
0& -{\bf 1} &0 \\
0&0& 4\tau^2\end{array}
\right)
+\frac{5\zeta(3)v^2}{64\pi^2 d^5}
\left(\begin{array}{ccc}
1&0&0\\
0& \frac{13}{5}{\bf 1} &0 \\
0&0& 4\tau^2\end{array}
\right)
\label{scal}
\end{eqnarray}
and for conformal coupling:
\begin{eqnarray}
\langle 0\vert \hat T_{MN}^S\vert 0\rangle_{ren}^{mov}\simeq
-\frac{3}{32\pi^2d^5}\left(\zeta(5)-\frac{\zeta(3)}{12}v^2\right)
\left(\begin{array}{ccc}
1&0&0\\
0& -{\bf 1} &0 \\
0&0& 4\tau^2\end{array}
\right)
\end{eqnarray}
The velocity contribution to the Casimir energy is  quadratic and 
positive in the 
non-relativisitic limit. In the following we will show that such contribution
is nothing but the energy density in the form of particles 
created from the vacuum by the moving boundaries. 

\section{Particle production}
Let us consider a quantized real scalar field:
\begin{eqnarray}
\phi(x)=\sum_{\vec k} \sum_n  (a_{\vec k n}u_{\vec k n}+a^\dagger_{\vec k n} 
u^*_{\vec k n})
\end{eqnarray}
The modes $u_{\vec k n}$ obeying the Klein-Gordon equation in Milne coordinates 
are given by:
\begin{eqnarray}
u^N_{\vec k n}(x)=Nf_{kn}(\tau)e^{i\vec k\vec x}\cos(\nu\chi),\;\;\; \mbox{Neumann}
\nonumber \\
u^D_{\vec k n}(x)=Nf_{kn}(\tau)e^{i\vec k\vec x}\sin(\nu\chi),\;\;\; \mbox{Dirichlet}
\end{eqnarray}
where the functions $f_{kn}(\tau)$ satisfy the equation:
\begin{eqnarray}
\left(\hat\partial_0\hat\partial_0+k^2+\frac{\nu^2}{\tau^2}
+\frac{1}{\tau}\hat\partial_0\right)f_{kn}(\tau)=0
\label{kg}
\end{eqnarray}
with $k^2=\vec k^2$, $\nu=n\pi/\chi_0$ and $n$ a positive integer number.
The solutions of equation (\ref{kg}) can be written in terms of Hankel functions
$f_{kn}=H_{i\nu}^{(1,2)}(k\tau)$ where the $1(2)$ functions behave
as positive frequency modes in the $\tau\rightarrow +(-)\infty$ limit.
The solutions are normalized with respect to the scalar product:
\begin{eqnarray}
(\phi_1,\phi_2)=-i\int_V d^3x\int_0^{\chi_0}
d\chi\,\tau
(\phi^*_1\hat\partial_0\phi_2 
-\hat\partial_0\phi^*_1\phi_2)
\end{eqnarray}
where we are working in a box of finite
volume $V$ in the ordinary three-space. We will take
the continuum limit at the end of the calculations. From the
expression above, we can obtain the normalization constant 
$N=\sqrt{\frac{\pi}{2 V\chi_0}}e^{-\nu\pi/2}$.

Because of the moving boundaries, a given solution which behaves as 
positive frequency at a given time will become a linear superposition
of positive and negative frequency modes at any later time. This implies
that the initial vacuum state will contain particles as seen by an
observer at a later time. Thus consider the solutions at a given time
for the static problem with a fixed brane separation $L$.
The positive frequency modes satisfying the different boundary
conditions are given by:
\begin{eqnarray}
v_{\vec k n}^N=Ne^{-i\omega t+i\vec k\vec x}\cos(qz),\;\;\;\mbox{Neumann} \nonumber \\
v_{\vec k n}^D=Ne^{-i\omega t+i\vec k\vec x}\sin(qz),\;\;\; \mbox{Dirichlet}
\end{eqnarray}
where $q=n\pi/L$,  $\omega^2=k^2+q^2$ and $N=(LV\omega)^{-1/2}$. 
These modes will define a different Fock space which can be built out of
the corresponding creation and annhilation operators which we will
denote by $b_{\vec k n}$, $b_{\vec k n}^\dagger$. Thus, for either
the real or the imaginary part of our scalar fields we have: 
\begin{eqnarray}
\phi(x)=\sum_{\vec k} \sum_n (b_{\vec k n}v_{\vec k n}
+b^\dagger_{\vec k n} v^*_{\vec k n})
\end{eqnarray}
The two sets of operators are related
in general by the following Bogolyubov transformation \cite{Birell}:
\begin{eqnarray}
b_{\vec k n}=
\sum_{\vec k'}\sum_{n'}\left({\alpha_{\vec k\vec k'nn'}} a_{\vec k'n'}
 +{\beta_{\vec k\vec k'nn'}^*}
a_{\vec k'n'}^{\dagger}\right)
\end{eqnarray}
The Bogolyubov
coefficients relating both solutions are given by:
$\beta_{\vec k\vec k',nn'}=(u_{\vec k n},v_{\vec k' n'}^*)$.
Following \cite{jauregui}, we are interested in the calculation
of the total number of particles created by the brane motion in the 
non-relativisitic limit $\chi_0\ll 1$, i.e. for large values of $\nu$. 
In that limit, we can use the large-order expansion of Hankel
functions in the calculation of the Bogolyubov
coefficients. In the present case it is enough with the
first term in the expansion which is given by \cite{Watson}:

\begin{eqnarray}
H_{i\nu}^{(1)}(k\tau)\simeq\sqrt{\frac{2}{\pi}}\frac{1}{\left(\nu^2+k^2\tau^2\right)^{1/4}}
\exp\left(\frac{\nu\pi}{2}+i(\nu^2+k^2\tau^2)^{1/2}-i\nu\,\mbox{arcsinh}
\left(\frac{\nu}{k\tau}\right)
-i\frac{\pi}{4}\right)
\end{eqnarray}

This expression is valid for large $\tau$, i.e. large 
brane separations and satisfies the equation of motion up to terms of 
$\Od(\nu^{-2})$.
In the non-relativistic limit, we have $t=x^0=\tau+\Od(\chi^2)$, 
$z=\tau\chi+\Od(\chi^3)$ and 
$v=\chi_0+\Od(\chi_0^3)$, so that we can write $L=\tau\chi_0+\Od(\chi_0^3)$
and $q=\nu/\tau+\Od(\chi_0^2)$ so that $\omega^2=k^2+\nu^2/\tau^2+\Od(\chi_0^2)$. 
Therefore for the Bogolyubov coefficients relating Neumann solutions we get:
\begin{eqnarray}
\beta_{\vec k\vec k' nn'}&\simeq&-i\int_V d^3x\int_0^L dz 
\cos\left(\frac{n'\pi z}{L}\right)
\cos\left(\frac{n\pi\chi}{\chi_0}\right)
\frac{e^{i(\vec k'-\vec k)\vec x}}{2\chi_0}
\frac{k^2\tau}{(\nu^2+k^2\tau^2)^{3/2}}e^{i\Phi(\nu,k,\tau)}
\nonumber \\ 
&=&-i\delta_{\vec k\vec k'}\delta_{nn'}\frac{k^2\tau^2}{4(\nu^2
+k^2\tau^2)^{3/2}}
e^{i\Phi(\nu,k,\tau)}+\Od(\chi_0^0)
\end{eqnarray}
where the phase is given by:
\begin{eqnarray}
\Phi(\nu,k,\tau)=2\omega\tau
-\nu\, \mbox{arcsinh}\left(\frac{\nu}{k\tau}\right)+\frac{\pi}{4}
\end{eqnarray}
and the same results is obtained for Dirichlet solutions, whereas mixed
Dirichlet-Neumann coefficients vanish.
Defining $\beta_{\vec k\vec k' nn'}
=\beta_{\vec k n}\delta_{kk'}\delta_{nn'}$,  
the total energy-density of 
the particles created for the scalar matter content in our model with
two complex scalar fields is:
\begin{eqnarray}
\rho_S= \frac{4}{L}\sum_{n=0}^{\infty}\int\frac{d^3k}{(2\pi)^3}\omega
\vert \beta_{\vec k n}\vert^2\simeq 
\frac{4}{L}\sum_{n=0}^{\infty}\int\frac{d^3k}{(2\pi)^3}
\frac{k^4\tau^4\omega}{16(\nu^2+k^2\tau^2)^3}
\end{eqnarray}
where we have already taken the continuum limit $V\rightarrow \infty$.
This is a divergent expression which can be renormalized using the zeta-function
method \cite{Hawking}. By means of a change of variables, we can write the above expression
as:
\begin{eqnarray}
\rho_S\simeq\frac{1}{8\pi^2\chi_0\tau^5}\sum_{n=0}^{\infty}\int_0^{\infty} dx
\frac{x^6}{(x^2+\nu^2)^{5/2}}=\frac{1}{16\pi^2\tau^5\chi_0}
B\left(\frac{7}{2},-1\right)\sum_{n=0}^{\infty}
\left(\frac{n\pi}{\chi_0}\right)^2 
\end{eqnarray}
where $B(p,q)=\Gamma(p)\Gamma(q)/\Gamma(p+q)$ is the beta function and 
$\sum_n n^{-s}=\zeta(s)$ is the Riemann zeta function. 
We can cancel the divergent contributions
from the integral and the series by using the analytical continuation of the
relation: 
$\zeta(s)=\Gamma\left(\frac{1-s}{2}\right)\pi^{s-1/2}\zeta(1-s)/
\Gamma\left(\frac{s}{2}\right)$, 
we obtain the following renormalized result:
\begin{eqnarray}
\rho_S\simeq\frac{5}{64\pi^2\tau^5\chi_0^5}\zeta(3)v^2
\end{eqnarray} 
which agrees with the first correction to the static Casimir energy in 
(\ref{scal}). We see that this velocity correction can be understood as
the contribution from the particles produced by the non-adiabatic motion
of the branes \cite{jauregui}.

\section{Fermionic contribution}
Although the fermionic Casimir effect has been considered previously 
in different contexts, only a few results in the case 
with moving boundaries can be found in the literature \cite{Mazzitelli}. 
In this section
we will consider the problem in our five dimensional model.
 
The fermionic energy-momentum tensor for massless fermions in five dimensions
can be written in Minkowski coordinates as:
\begin{eqnarray}
T_{AB}^F&=&\frac{i}{4}\left(\bar\psi\gamma_A \partial_B\psi
+\bar\psi\gamma_B \partial_A\psi 
-\partial_A\bar \psi\gamma_B\psi- \partial_B\bar \psi\gamma_A\psi \right)
\label{TF}
\end{eqnarray} 
Following similar steps to the scalar case, we will
compute
the corresponding vacuum expectation value, in the case in which 
the two branes are fixed, by means of the image method. 
The fermionic Hadamard function
 can be written as:
\begin{eqnarray}
{\cal D}^{(1)}_{ab}(x,y)&=&\langle 0\vert 
\bar\psi_b(y)\psi_a(x)-\bar\psi_b(x)\psi_a(y)\vert 0\rangle
\end{eqnarray}
where $a,b$ are spinor indices.
From (\ref{TF}), the vacuum expectation value of $T_{AB}^F$  can be
written in Minkowski
space-time in terms of the Hadamard function as:
\begin{eqnarray}
\langle 0\vert T_{AB}^F\vert 0\rangle
=\lim_{x'\rightarrow x,\; y'\rightarrow x}
\frac{i}{4}\left((\gamma_A)_{ba} \partial_B^{x'}
+(\gamma_B)_{ba} \partial_A^{x'}\right)
{\cal D}^{(1)}_{ab}(x',y')
\label{tensorf}
\end{eqnarray}

In the simplest case  with no boundary conditions
and in Minkowski coordinates, the Hadamard function  is given in terms
of the scalar function (\ref{sfree}) as 
${\cal D}^{(1)}_{ab}(x,y)=-i\gamma^A_{ab}\partial_A^x D^{(1)}(x,y)$. 
In the case with fixed boundaries, the different fermionic 
components satisfy (\ref{boundary}) and 
the  Hadamard function obeying these boundary conditions
can be written as:
\begin{eqnarray}
{\cal D}^{(1)}(x,y)&=&P_L{\cal D}^{(1)}_{DD}(x,y)P_R+
P_R{\cal D}^{(1)}_{NN}(x,y)P_L\nonumber \\
&+&
P_L{\cal D}^{(1)}_{DN}(x,y)P_L+
P_R{\cal D}^{(1)}_{ND}(x,y)P_R
\label{hada}
\end{eqnarray}
With these relations, 
we obtain from (\ref{tensorf}) 
the following expression for the energy-momentum tensor 
between the fixed branes:
\begin{eqnarray}
\langle 0\vert  T_{AB}^F\vert 0\rangle
&=&\lim_{x'\rightarrow x,\; y'\rightarrow x} 
\frac{1}{4}\tr \left(\partial_B^{x'} {\cal D}_{DD}^{(1)}(x',y')
P_R\gamma_A P_L+\partial_B^{x'} {\cal D}_{NN}^{(1)}(x',y')
P_L\gamma_A P_R\right.\nonumber \\
&+&\left.\partial_B^{x'} {\cal D}_{DN}^{(1)}(x',y')
P_L\gamma_A P_L+\nabla_B^{x'} {\cal D}_{ND}^{(1)}(x',y')
P_R\gamma_A P_R\right)+(A\leftrightarrow B)\nonumber \\
\label{tmnferm}
\end{eqnarray} 

In order to calculate the explicit expression of the different components
of the Hadamard function in (\ref{hada}), 
we need to know the mode expansion of the
Dirac field:
\begin{eqnarray}
\psi(x)=\sum_{\vec k n \alpha}(b_{\vec k n\alpha}v^+_{\vec k n\alpha}(x)
+d_{\vec k n\alpha}^\dagger v^-_{\vec k n\alpha}(x))
\label{modexp}
\end{eqnarray}
where we have labelled the states by $\alpha=\pm 1/2$, according to 
the value of the spin projection along the ordinary three-momentum
on the brane. The positive and negative frequency solutions of the
Dirac equation in Minkowski coordinates are given by:
\begin{eqnarray}
v_{\vec k n\alpha}^\pm(x)=\frac{1}{\sqrt{2LV\omega(\omega\pm q)}}
e^{\mp i\omega t}e^{i\vec k\vec x}
\left(
\begin{array}{c}
\sin(qz)U_\alpha\\
\cos(qz)U_\alpha
\end{array}
\right)
\end{eqnarray}
where the two-component spinors with the appropriate helicities are given by: 
$U_{1/2}^T=(1,0)$ and $U_{-1/2}^T=(0,1)$ and the component of the momentum in
the fifth dimension $q$ is defined as in the scalar case with fixed
boundaries. 
The modes are normalized with
respect to the scalar product:
\begin{eqnarray}
(\psi_1,\psi_2)=\int d^4x \sqrt{g_0}\bar\psi_1\hat\gamma_0\psi_2
\label{scalarprod}
\end{eqnarray}
where $g_0$ denotes the determinant of the induced metric on the
hypersurfaces of constant time.
In the calculation of the energy-momentum tensor in (\ref{tmnferm}) 
only derivatives of the Green functions are relevant, thus, making use of the
the equations of motion, $\pabar^x{\cal D}^{(1)}(x,y)=0$ and acting 
with the projectors we get:
\begin{eqnarray}
\gamma^4\partial_4^x P_R{\cal D}_{ND}^{(1)}(x,y) P_R
=-\gamma^\mu\partial_\mu^x P_L{\cal D}_{DD}^{(1)}(x,y)P_R
\label{truco}
\end{eqnarray}
and a similar expression is obtained for the $NN$ function.
Tracelessness and conservation determine a diagonal
form of the energy-momentum tensor in the Minkowski background.
Accordingly the knowledge of the  $DD$ and $NN$ functions
together with Eq. (\ref{truco}) is sufficient to compute the
complete tensor. Inserting the
explicit mode expansion (\ref{modexp}) into 
(\ref{hada}), it can be shown that such functions are given by:
\begin{eqnarray}
P_L{\cal D}^{(1)}_{DD}(x,y)P_R=-iP_L\pabar^x D_{D}^{(1)}(x,y)P_R
\end{eqnarray} 
where $D_{D}^{(1)}(x,y)$ is the scalar
function given in (\ref{Di}) and a similar expression is obtained 
for the $NN$ function. 
From (\ref{tmnferm}) we finally get:
\begin{eqnarray}
\langle 0\vert  T_{AB}^F\vert 0\rangle_{ren}^{static}=
2\lim_{x'\rightarrow x,\; y'\rightarrow x}
\sum_{n=-\infty}^{\infty}
\partial^{x'}_A\partial^{x'}_A D^{(1)}(x',y'_{n+})\delta_{AB}=
\frac{3\;\zeta(5)}{32\pi^2  L^5}
\left(\begin{array}{ccc}
1&0&0\\
0& -{\bf 1} &0 \\
0&0& 4\end{array}
\right)
\end{eqnarray}
which agrees with the static scalar case (\ref{scalarest}) but with
opposite sign. Accordingly, and as expected from supersymmetry, for
static branes the total energy-momentum vanishes.

The case in which one of the branes is moving cannot be computed easily
with the image method, so that in the following we will use the Bogolyubov
transformations technique. Let us then consider again
Milne coordinates. 
We introduce the following expression for the 
vielbein $e^A_{\;M}$ and the inverse vielbein $\hat e^M_{\;A}$
corresponding to the Milne metric in (\ref{Milne}): 
$\hat e^\mu_{\;\nu}=e^\mu_{\;\nu}=\delta^\mu_{\;\nu}$ with $\mu=0,..,3$, 
and $e^4_{\;4}=\tau$,
$\hat e^4_{\;4}=1/\tau$. The curved gamma matrices are defined as
usual by 
$\hat \gamma^M=\hat e^M_{\;A}\gamma^A$. The Dirac equation in
these coordinates reads:
\begin{eqnarray}
i\hat{\Dbar} u=\left(i\gamma^\mu\hat\partial_\mu+\frac{i}{\tau}\gamma^4
\hat\partial_4+\frac{i}{2\tau}\gamma^0\right)u=0
\label{Dirac}
\end{eqnarray}
We look for solutions in the form $u_{\vec k n \alpha}^\pm=
i\hat{\Dbar} \hat\psi_{\vec k n \alpha}^\pm$ with:
\begin{eqnarray}
\hat\psi_{\vec k n \alpha}^\pm(x)=N
f_{\vec k n}^\pm(\tau)e^{i\vec k\vec x}
\left(
\begin{array}{c}
\sin(\nu \chi)U_\alpha\\
\cos(\nu \chi)U_\alpha
\end{array}
\right)
\end{eqnarray}
where $\pm$ index denotes positive or negative frequency and
again the fifth component of the momentum $\nu$ is defined as in the
scalar case with moving boundaries.
With these definitions it can be seen that the spinor 
$u_{\vec k n \alpha}^\pm$ 
obeys the
correct boundary conditions, i.e.:
\begin{eqnarray}
P_L\,u_{\vec k n \alpha}^\pm\vert_{\chi=0,\chi_0}&=&0,\nonumber \\
\partial_\chi P_R\,u_{\vec k n \alpha}^\pm\vert_{\chi=0,\chi_0}&=&0
\label{fermbound}
\end{eqnarray}
The function $f_{\vec k n}^\pm(\tau)$ satisfies the equation:
\begin{eqnarray}
{f_{\vec k n}^\pm}''(\tau)+\left(k^2+\frac{\nu^2}{\tau^2}
-\frac{1}{4\tau^2}+\frac{i\nu}{\tau^2}\right)f_{\vec k n}^\pm(\tau)
+\frac{1}{\tau}{f_{\vec k n}^\pm}'(\tau)=0
\end{eqnarray}
The corresponding solutions are given by Hankel functions 
$f_{\vec k n}^+(\tau)=H^{(1)}_{i\nu-1/2}(k\tau)$ and 
$f_{\vec k n}^-(\tau)=H^{(2)}_{i\nu-1/2}(k\tau)$. 
In the non-relativistic limit $\nu\gg 1$ the
asymptotic expansion of the positive frequency 
Hankel function reads \cite{Watson}:

\begin{eqnarray}
H^{(1)}_{i\nu-1/2}(k\tau)&\simeq&\sqrt{\frac{2}{\pi}}\frac{1}{(k^2\tau^2+\nu^2)^{1/4}}
\left(1-\frac{i\nu}{4(k^2\tau^2+\nu^2)}\right)\nonumber \\
&\times&\exp\left(i(k^2\tau^2+\nu^2)^{1/2}+\frac{\nu\pi}{2}-\left(i\nu-\frac{1}{2}
\right)\mbox{arcsinh}
\left(\frac{\nu}{k\tau}\right)\right)
\end{eqnarray}

This expansion satisfies Eq.(\ref{Dirac}) up to $\Od(\nu^{-2})$.
The normalization constant can be obtained from the scalar product
(\ref{scalarprod}) and is given by: 
$N=\frac{1}{2}e^{-\pi\nu/2}\sqrt{\frac{\pi}{kV\chi_0}}$. Following similar
steps to the scalar case, we calculate the number of fermions produced by means
of the Bogolyubov coeffcient $\beta_{\vec k\vec k',nn',\alpha\alpha'}=
(u_{\vec kn\alpha}^+,v_{\vec k'n'\alpha'}^-)$. After a lengthy calculation
we obtain $\beta_{\vec k\vec k',nn',\alpha\alpha'}=
\beta_{\vec k n \alpha}
\delta_{\vec k\vec k'}\delta_{nn'}
\delta_{\alpha\alpha'}$ with $\beta_{\vec k n \alpha}=0$ to the
lowest order i.e. $\Od(\nu^0)$ and
\begin{eqnarray}
\beta_{\vec k n \alpha}&\simeq&\frac{i}{4k\nu\left(k^2
+\frac{\nu^2}{\tau^2}\right)^{3/2}}
\left(\frac{\nu}{\tau}\left(k^2+\frac{\nu^2}{\tau^2}\right)^{3/2}
+\frac{\nu^2}{\tau^2}\left(k^2+\frac{\nu^2}{\tau^2}\right)+
\frac{\nu^2k^2}{\tau^2} \right)\nonumber \\
&\times&
\exp\left(2i\omega\tau-i\nu\, \mbox{arcsinh}\left(\frac{\nu}{k\tau}\right)
+i\frac{\pi}{4}\right)
\end{eqnarray}
including the first order ($\Od(\nu^{-1})$) correction.

The total energy density to this order is then given by:
\begin{eqnarray}
\rho_F=\frac{2}{L}\sum_{\alpha=\pm 1/2}
\sum_{n=0}^{\infty}\int \frac{d^3k}{(2\pi)^3}\omega \vert 
\beta_{\vec k n\alpha}\vert^2
\end{eqnarray}
where the factor of two comes from the fact that antifermions are 
produced in the same amount as fermions. Using again zeta-function 
regularization
to perform the integrals we obtain:
\begin{eqnarray}
\rho_F&\simeq&\frac{\pi^2}{16\tau^4\chi_0^2L}
\left(B\left(\frac{1}{2},-1\right)+B\left(\frac{1}{2},0\right)+
B\left(\frac{5}{2},0\right)+B\left(\frac{1}{2},-\frac{1}{2}\right)\right.
\nonumber \\
&+&\left.B\left(\frac{3}{2},-\frac{1}{2}\right)+B\left(\frac{3}{2},0\right)\right)
\sum_{n=0}^\infty n^{2}=-\frac{9}{64\pi^2\tau^5\chi_0^3}\zeta(3)
\end{eqnarray}

Finally combining this result with the static contribution in (\ref{scalarest}) 
and taking into account that for massless fermions the theory is conformally
invariant, the total result for the renormalized energy-momentum tensor
up to second order in the velocity is:
\begin{eqnarray}
\langle 0\vert \hat T_{MN}^F\vert 0\rangle_{ren}^{mov}\simeq 
\frac{3}{32\pi^2d^5}\left(\zeta(5)-\frac{3\zeta(3)}{2}v^2\right)
\left(\begin{array}{ccc}
1&0&0\\
0& -{\bf 1} &0 \\
0&0& 4\tau^2\end{array}
\right)
\end{eqnarray}

We can compute the total result including minimal  scalar
and fermion contributions, it is given by:
\begin{eqnarray}
\langle 0\vert \hat T_{MN}^{tot}\vert 0\rangle^{mov}=\langle 0\vert 
\hat T_{MN}^S\vert 0\rangle_{ren}^{mov}
+\langle 0\vert \hat T_{MN}^F\vert 0\rangle_{ren}^{mov}\simeq
-\frac{\zeta(3)v^2}{16\pi^2d^5}
\left(\begin{array}{ccc}
1&0&0\\
0& -\frac{11}{2}\,{\bf 1} &0 \\
0&0& 4\tau^2\end{array}
\right)
\end{eqnarray}
For conformal coupling we get:
\begin{eqnarray}
\langle 0\vert \hat T_{MN}^{tot}\vert 0\rangle^{mov}\simeq
-\frac{17\,\zeta(3)v^2}{128\pi^2d^5}
\left(\begin{array}{ccc}
1&0&0\\
0& -\,{\bf 1} &0 \\
0&0& 4\tau^2\end{array}
\right)
\end{eqnarray}

We see that the result is non-vanishing for $v \neq 0$, i.e. a Casimir
effect arises between the branes even though we are using the same kind of
boundary conditions which preserved supersymmetry  in the static case.

\section{Conclusions and discussion}

In this work we have considered the dynamical Casimir effect in a five
dimensional global supersymmetric model between two moving boundary branes.
Taking the same kind of boundary conditions
as in the static case, the results 
with moving boundaries are however completely different. In fact a non-vanishing
negative Casimir energy is generated both for minimal and
conformally coupled scalars. This induces a new
type of velocity dependent potential between the branes and
the breaking of supersymmetry due to the branes motion. 

The presence of this additional energy density  could have
interesting cosmological effects. Thus for instance, in the context
of the ekpyrotic (cyclic) model \cite{TS}, the universe contracts
to a singularity and then reexpands. However 
it was shown in \cite{Linde} 
that  ultra-relativistic particles produced 
near the singularity could disrupt the
cyclic evolution provided their energy density reaches the Planckian scale.  
We have shown that the velocity contribution to the 
Casimir effect can be interpreted as particle production in the
Minkowski vacuum and in fact
the energy density is expected to be very high in the limit of small
brane separation.
The stability of this kind of
orbifold singularities has been also studied in \cite{Polchinski}.
According to their conclusions, the introduction of a single particle
between the branes causes the collapse of the universe 
into a curvature singularity.

Concerning the problem of particle production in a Milne 
metric, in a recent paper \cite{Turok}, it has been shown that by  matching 
modes through the singularity for a free theory, it is possible to
find an appropriate vacuum  state in which particles are not produced. 
However, when including
time-dependent interactions,  particles are generically produced.
We understand that this should be the case in the presence of the 
moving boundaries, implying that the energy-momentum
tensor calculated in this work is physical and should backreact on the 
space-time geometry. (See also \cite{Seiberg} for a discussion about
the backreaction problem.)

\section*{Acknowledgments}

I am grateful to A. Linde for useful discussions and
important suggestions and to C. Herdeiro, S. Hirano
and  L. Kofman for useful comments and discussions. 
This work has been partially supported by the CICYT (Spain) project
 FPA2000-0956. The author  also acknowledges support from 
the Universidad Complutense del Amo Program.

\end{document}